\documentstyle[prl,epsfig,aps]{revtex}
\begin{document}

\twocolumn[\hsize\textwidth\columnwidth\hsize\csname@twocolumnfalse\endcsname 
 
\title{Edwards' measures: a thermodynamic construction  for dense
granular media and glasses} 
 
\author{Alain Barrat$^1$, Jorge Kurchan$^2$,  
Vittorio Loreto$^3$  and Mauro Sellitto$^4$ }  
  
\address{   
$^1$ Laboratoire de Physique Th{\'e}orique  
\cite{umr}, B{\^a}timent 210, Universit{\'e} 
de Paris-Sud, 91405 Orsay Cedex, France \\ 
$^2$ P.M.M.H. Ecole Sup\'erieure de Physique et Chimie 
Industrielles, 10 rue Vauquelin 75231 Paris, France \\ 
$^3$  Universit\`a degli Studi di Roma ``La Sapienza'', 
Dipartimento di Fisica and INFM Unit\`a di Roma 1, \\
P.le A. Moro 5, 00185 Rome, Italy\\
$^4$ Laboratoire de Physique 
             de l'\'Ecole Normale Sup\'erieure de Lyon,         
  46 All\'ee d'Italie, 69007 Lyon, France. 
} 
 
\date{\today} 
 
\maketitle 
\begin{abstract} 

We present numerical support for the hypothesis that
macroscopic observables of dense granular media
can be evaluated from averages over {\em typical} blocked configurations:
we construct the corresponding measure for a class of
finite-dimensional systems and compare its predictions for various observables
with the outcome of the out of equilibrium dynamics at large times.
We discuss in detail the connection with the effective temperatures that
appear in out of equilibrium glass theories, as well as
the relation between our computation and those
based on `inherent structure' arguments.
A short version of this work has appeared in Phys. Rev. Lett.
{\bf 85}, 5034 (2000), cond-mat/0006140.

\end{abstract} 
\twocolumn  
\vskip .5pc] 
\narrowtext 
 
\section{introduction}

Granular systems \cite{Nagel,Nagel2} involve many particles, 
so there is a strong
motivation to treat them with thermodynamic methods.
This approach is justified when  
one is able to identify a distribution that is 
left invariant by the dynamics ({\em e.g.} the microcanonical ensemble), and 
then assume that this distribution will be reached by the system, under  
suitable conditions of 'ergodicity'. 
Unfortunately, because energy is lost through internal friction, and 
gained by a non-thermal source such as tapping or shearing, 
the dynamical equations do not leave the microcanonical or any 
other known ensemble invariant. Moreover, just as in the case of aging
glasses, the compaction 
dynamics  does not approach any  
stationary state on experimental time scales. 

Consider  a compaction experiment, in which we subject a granular
system to gentle, periodic tapping. To keep the discussion simple,
we can assume that
there is no gravity, and that there is a piston applying a constant
pressure on the surface. The system compactifies very slowly
\cite{Knight}, in practice never reaching the most dense,
optimal packing. 
At a given long time, when the system has density $\rho(t)$, we may
wish to measure for example the fraction of grains that are at relative
distance $r$: the structure factor.  
This quantity being averaged over all particles, one can expect it 
to be a reproducible observable.
However, there is in principle no method to calculate the structure
factor other than solving the dynamics.

Some years ago Edwards \cite{Sam,anita,Repo}
proposed that one could reproduce the observables  attained  
dynamically  by calculating the value they take    in
the usual equilibrium distribution at the corresponding volume,  
energy, etc. {\em but restricting the sum to the `blocked' configurations}  
defined as those in which every grain is unable to move.
In the case of the previous paragraph, we would compute the structure
factor in {\em all the possible} blocked
configurations of density $\rho(t)$, and calculate the average. Thus,
the only input from dynamics would be   $\rho(t)$, apart from which
the calculation is based on a statistical ensemble.

This `Edwards ensemble'  
 leads immediately to the definition of an entropy  (in the glass  
literature a `complexity') $S_{edw}$, given by  the logarithm  
of the number of blocked configurations of given volume, energy, etc., 
and its corresponding  density $s_{edw}\equiv S_{edw}/N$. 
Associated with this entropy are state variables such as 
`compactivity' $X_{edw}^{-1}=\frac{\partial}{\partial V}S_{edw}(V)$ and 
`temperature' $T_{edw}^{-1}=\frac{\partial}{\partial E}S_{edw}(E)$. 

Recent developments in glass theory, especially those related to their
out of equilibrium dynamics, have come to clarify and support such a
hypothesis --- at least within mean-field models (see below).
The present paper addresses the  natural question of 
 whether Edwards' measure gives
 good results for the compaction of  finite
 dimensional, non mean-field models. 
The result is that there is a class of models for which 
this is the case. 
A shorter version of this work has been published in \cite{bklm}.

The article is organized as follows: we first discuss 
the nature of  the assumptions (Subsection I-A), and the 
evidence in support coming from mean-field models (Subsection I-B).
In subsection I-C we discuss in detail the checks already made with
Lennard-Jones glasses, in the context of the so-called `inherent
structures',
and their relation with the present approach. 

In sections II and III we treat two  finite-dimensional models 
which  reproduce
many of the features of glasses and granular media, namely
the Kob-Andersen (KA)\cite{KoAn} and Tetris \cite{prltetris} 
models. We devise a method
to count and calculate averages over the blocked configurations, 
explicitly constructing in this way Edwards' 
measure. We compare expectation values
thus obtained with equilibrium values and with the outcome 
of slow, aging dynamics and find very good agreement between
the predictions of Edwards' measure and aging dynamics.
 
In order to show that this agreement does not hold 
for all forms of slow dynamics,
we repeat the procedure in  section IV for another model 
exhibiting slow, logarithmic relaxations, the random field Ising model
(RFIM) \cite{Nattermann}. We conclude 
with a discussion of our results in section V.

\subsection{The Assumption} 

 One possibility of making an assumption \`a la Edwards would be
to consider a fast quench, and then propose that the configuration
reached has the macroscopic properties of the typical blocked
configurations. This would imply that the system stops at a density
for which the number of blocked configurations is maximal.
We do not follow this path,
as we will give sufficient evidence that generically the vast majority
of the blocked configurations are much less compact than the one
reached dynamically, even after abrupt quenches.

Our strategy here is instead to quench the system to a situation of
very weak but non-zero tapping, shearing or thermal agitation.
In this way, the system keeps compactifying, albeit at a very slow
rate.
In this context, we consider a flat measure over blocked 
configurations {\em conditioned to having the energy and/or density
of the dynamical situation we wish to reproduce}.
This means that we have given up trying to predict the dynamical energy 
or density by methods other than the dynamics itself.

That configurations with low mobility should be relevant in a jammed 
situation is rather evident, the strong hypothesis here is that the 
configurations reached dynamically are {\em the typical ones} of given
energy and  density.
Had we restricted averages to blocked configurations having {\em all}
macroscopic observables coinciding with the dynamical ones, the
construction would exactly, and trivially, reproduce the dynamic
results. The fact that conditioning averages to the observed
energy and density suffices to give, at least as an approximation, 
other dynamical observables is highly non-trivial.

At this point it is important 
to warn the reader about a  misconception. It goes like this:
{\em The system has at every energy and density 
many blocked configurations. Now, we
know that in systems with many minima (for example from mean-field
glasses) all the minima of 
 given energy tend to have the same values for macroscopic observables.
Hence, it is natural to assume that their basins of attraction are
themselves
also the same, and hence Edwards' `flat average' assumption is justified.}

To see the danger of such a reasoning, 
let us paraphrase it in another context,
in which it is clear that the conclusion is generically erroneous:
consider a {\em driven,
macroscopic  system} \cite{Ott}
(e.g.  fully developed turbulence)
thermostated  at energy $E$. By the same token, we would say: 
{\em The system is restricted to move in 
the energy $=E$ shell. Now, we
know that almost all points in an energy shell of a macroscopic system have
 in the thermodynamic limit the same values of  macroscopic observables
(we disregard symmetry breaking cases).
Hence, it is natural to suppose  that the dynamic stationary measure
is also
the same in all points: therefore the stationary 
distribution is flat, i.e. microcanonical.}

This is of course  wrong: we know that generically the
stationary  measure of a driven, thermostated system is dominated by
an ensemble of zero volume within the energy shell.
Almost all  points in the energy shell do have the same
weight, {\em but this weight is zero}.
In order that the stationary measure coincides with the microcanonical measure
we need some strongly specific properties for the dynamics: such is the case
of chaotic {\em Hamiltonian} dynamics.
The same can be said about Edwards' measure: that all blocked
configurations
of a given energy have the same basin of attraction may be quite generally
true, but in order for Edwards' measure to be relevant the combined
basin of attraction of
the typical configurations should not vanish!

Starting from a random configuration the probability of falling into a
basin is proportional to its volume, so with a quench we are not
sampling a typical basin: rare basins (exponentially smaller in
quantity) of large (exponential) volume may dominate ---
and this is generally true for a  quench  from  equilibrium
at any temperature.
Using Edwards' measure is justified when for some reason one can
consider that typical basins of a given level
 are also typically accessed: a very strong
assumption that is generally not valid.
The reason this hypothesis (if true) is useful is that one can 
construct averages over configurations defined by a local property
(being blocked) without having to know the basin of attraction,
which involves  solving the dynamics.

There is still however a puzzling  question:
We are using blocked configurations  as a distribution
 for the dynamic situation. However, this seems odd, since we know
that
neither a relaxing nor a gently driven system  will  stay in one of those
configurations! Here the example of the stationary 
 measure in dynamic systems is
also instructive: we know that such a measure can be constructed by
considering only the periodic trajectories,
although the probability that the system is
{\em in} a periodic trajectory is strictly zero. Somehow, these trajectories
form a `skeleton' of the true distribution --- and  such would also
be the role of the blocked configurations in Edwards' measure.

Moreover, we will check that configurations with a small, though non-zero
fraction of mobile particles yield the same statistics.

\subsection{Solvable models}

As mentioned above, the fact that dynamically accessed 
 blocked configurations are the typical ones does not follow from 
 any general principle that we know, and, as we shall see 
below,  is indeed not always true. 
 
In order to progress, one can exploit the analogy  
between the settling of grains and powders,   
and the aging of glassy systems~\cite{Struik} since 
in both cases, 
the system remains out of equilibrium on all accessible time-scales, 
and displays very slow relaxations.  
 
In the late eighties, Kirkpatrick et al.~\cite{KTW,KTW2} recognized that 
a class of mean-field models contains, although in a rather 
schematic way, the essentials of glassy phenomena. 
When the aging dynamics of these systems was solved analytically, 
a feature that emerged was the existence of a temperature $T_{dyn}$ 
for {\em all} the slow modes (corresponding to structural rearrangements) 
\cite{review,Cukupe}. 

For the purposes of this paper,  $T_{dyn}$ can be defined by comparing 
the random diffusion and the mobility between two widely 
separated times $t$ and $t_w$ of any particle or tracer 
in the aging glass.  
Surprisingly, one finds  in all cases an  Einstein 
relation  $\left\langle (r(t)-r(t_w))^2 \right\rangle 
 = T_{dyn}\frac{\delta \left\langle r(t) -r(t_w) \right\rangle }{\delta f}$, 
where $r$ is the position of the particle and
$f$ is a constant perturbing field, and the brackets denote
average over realizations.
While in an equilibrium system the fluctuation-dissipation theorem  
guarantees that the role of $T_{dyn}$ is played by the  
thermodynamic temperature, 
the appearance of such a quantity out of equilibrium is by no  
means obvious. $ T_{dyn}$ is different from 
the external temperature, but it can be shown to have 
 all other properties defining a true temperature~\cite{Cukupe}. 
 
As it turned out, despite its very different origin, 
this temperature matches exactly Edwards' ideas.
One can identify in mean-field models all  the energy minima (the blocked
configurations in a gradient descent dynamics), and calculate   
$1/T_{edw}$ as the derivative of the logarithm of their number with respect
to energy. An explicit computation shows that $T_{edw}$ coincides with
$T_{dyn}$ obtained from the out of equilibrium dynamics  of the
 models aging in contact  
with an almost zero temperature 
bath~\cite{remi,jamming,Theo,Frvi,Felix,biroli}.  
Moreover, given the energy $E(t)$ at long times, the value of any other 
macroscopic observable is also given by  the flat average over all blocked 
configurations of energy $E(t)$. 
Within the same approximation, one can also treat systems that like 
granular matter present a non-linear friction and different kinds of energy 
input, and the conclusions remain the same~\cite{jorge-trieste}
despite the fact that there is no thermal bath temperature.  
 
Edwards' scenario then  
happens to be  correct within mean-field schemes and for  
very weak  vibration or forcing. The problem that remains is  
to what extent it carries through to more realistic models.  
 In this direction, there have been  recently  studies~\cite{KoScTa}  
of Lennard-Jones glass formers from the perspective of the so-called  
`inherent structures' \cite{inherent_equilibrium}, 
which suggest  that whatever the measure for
the slow dynamics, it is not sensitive to the details of the thermal
 history --see next subsection. 

The path we  follow here \cite{bklm} is  to construct the Edwards 
measure explicitly in the case of  
representative (non mean-field) systems, together with the corresponding 
entropy and expectation values of observables. 
We thus obtain results that are clearly  different  
from the equilibrium ones, 
and we can compare both sets  with those  
of the  irreversible compaction dynamics.  
 
\subsection{Differences and similarities with 
approaches based on `inherent structures'}

Let us discuss in detail the relation
between the present approach \cite{bklm} with the one followed by Kob
et al.\cite{KoScTa} with Lennard-Jones glasses,
 later applied in  the granular matter context in  
\cite{CoNi}.

One can describe the work of Kob et al. as follows:
starting from an equilibrated system  at  temperature
 $T$ above the glass transition,  the system is first
quenched to a temperature $T_f$ below the glass transition, where 
the system  spends  aging a time  $t_w$, after which it is quenched again
to zero temperature (In some of 
the procedures there is no intermediate stop: $t_w=0$).
The dynamics of the final quench being  at zero temperature,
 the system eventually lands in
 a blocked configuration of energy $E$.
 For each cooling protocol parameterized by $(T,T_f,t_w)$
 the final energy $E_{T,T_f,t_w}$ is  a  reproducible quantity in the
 thermodynamic limit.

 Suppose now we classify all
thermal histories according to the  energy $E$ of the blocked 
configuration reached at the end. 
Kob et al. then ask the following question: are all other macroscopic
 observables
fully determined by $E$, or, otherwise stated, 
 is the effect of the whole history $(T,T_f,t_w)$ completely encoded in $E$?
For the macroscopic observables they considered,
their answer is within numerical precision   affirmative.
(In their work, they chose as macroscopic observables
the spectrum of the energy Hessian, instead of the structure factor as
we do here). 

We can now discuss the relation of their approach and
  the present paper. On the one hand, 
 because there is no direct sampling of typical configurations
of energy $E$, but a comparison of configurations reached after
different
histories,  the procedure of Ref. \cite{KoScTa} does not address the
question of what  the
actual distribution is for given $E$ (unless, of course, one
makes extra assumptions). 
One could imagine a situation in which a small subset of blocked configurations
of given energy contributes to the measure because they have a larger basin
of attraction for every history. Indeed, we shall see later that the 
dynamics of the RFIM, while passing the test of Kob et al., is not
well reproduced by a flat Edwards' measure.
What Ref.\cite{KoScTa}
does suggest is that whatever the measure, it is insensitive
to the details of the thermal history, which only has the effect of
specifying $E$ --- two thermal histories finishing in the same energy $E$
yield the same values for all the other macroscopic observables.
The approaches are clearly complementary: the suggestion of the
present work that Edwards' measure gives good results for a slow
compaction would be of little use without  the insensitivity of the measure
 on the history suggested in Ref. \cite{KoScTa}. 

Let us add that Kob et al. define a temperature $T^*$ for a process
$(T,T_f,t_w)$ by  demanding that $E$ obtained by a direct quench
from $T^*$ to zero temperature be equal to  $E_{T,T_f,t_w}$.
This temperature is not equal to  (but may be an approximation of)
the Edwards temperature which we calculate below.

\section{KA model}

The first model we consider is the so-called Kob-Andersen (KA)  
model~\cite{KoAn} that was first studied in the context
of Mode-Coupling theories \cite{gotze} as a finite dimensional model
exhibiting a divergence of the relaxation time at a finite
value of the control parameter (here the density); this divergence is due
to the presence in this model of the formation of ``cages'' around particles
at high density (the model was indeed devised to reproduce the cage effect
existing in super-cooled liquids).

Though very schematic, it has then been shown to reproduce rather well  
several aspects of glasses~\cite{KuPeSe}, like
the aging behaviour with violation of FDT \cite{Se}, and of granular  
compaction~\cite{SeAr}. 

The simplicity of its definition and the fact that it is non mean-field
makes it a very good candidate to test Edwards' ideas: in fact, the 
triviality of its Gibbs measure will allow us to compare the numerical
data obtained for the dynamics and for Edwards' measure with the analytic
results for equilibrium.

\subsection{Definition}

The model is defined as a lattice gas on a three dimensional lattice,
with at most one particle per site. The dynamical rule is as follows:
a particle can move to a neighboring empty site, only if it  
has strictly less than $m$ neighbours in the initial and in the final 
position. 
Following~\cite{KoAn}, we take $m=4$: this ensures that the system is
still ergodic at low densities, while displaying a sharp increase
in relaxation times
 at a density well below $1$.
The dynamic rule guarantees that the equilibrium distribution is trivially  
simple since all the configurations of a given density are equally probable:
the Hamiltonian is just $0$ since no static interaction exists. 

In order to mimic a compaction (or aging) process 
without gravity, we simulate a `piston'  by creating and  
destroying particles only on the topmost layer (of a cubic lattice
of linear size $L$) with a  chemical  
potential $\mu$~\cite{KuPeSe}. More precisely, each Monte-Carlo
sweep is divided in the following steps:
(i) for each of the site of the topmost layer, add a particle if the
site is empty, and, if it is occupied, withdraw the particle
with probability $\exp(-\beta \mu)$.
(ii) try to move each particle, in random order, according to the
dynamical rule.

\subsection{Gibbs measure}

Since the Hamiltonian is 0, the equilibrium (or Gibbs)
measure corresponds simply
to a flat measure over all configurations, without taking into account
the dynamical constraint. Therefore, 
the relation between density and
chemical potential is
\begin{equation}
  \rho = \frac{1}{1+\exp(- \beta \mu)} \ ,
  \label{eq:rhoeq}
\end{equation}
and the exact equilibrium entropy 
density per particle reads 
\begin{eqnarray}
  s_{equil}(\rho) &=& -\rho \ln \rho - (1-\rho)\ln(1-\rho) \,.
  \label{eq:s_equil}
\end{eqnarray} 
with in particular
\begin{eqnarray}
  \frac{ds_{equil}}{d\rho} &=& - \beta \mu \,.
\label{eq:mu_eq}
\end{eqnarray}
In this model, the temperature $1/\beta$ is irrelevant since it
appears only as a factor of the chemical potential 
and we can set it to one throughout. 

Besides, the equilibrium structure factor defined as
the probability that two sites at distance $r$ are both occupied 
is easily seen to be 
a constant 
\begin{eqnarray}
  g_{equil}(r) &=& \rho^2 \, \;\;\; ; \;\;\; r>0
  \label{eq:g_equil}
\end{eqnarray}
No correlations appear since the configurations are generated by putting 
particles at random on the lattice.
It will therefore be easy, as already mentioned, to compare small deviations  
from $g_{equil}(r)$, a  notoriously difficult task to do in 
glassy systems. Note that it is also easy to numerically sample
Gibbs' measure at any given density, 
by simply generating at random configurations with fixed
number of particles.

\subsection{Non-equilibrium dynamics}

The previously described Monte-Carlo procedure allows to produce
equilibrium configurations, even if the dynamical constraint is enforced,
as long as $\mu$ is low enough.
However, at densities close to $\rho_{\rm g}$ ($\simeq 0.88$), the particle
diffusion becomes extremely slow due to the kinetic constraints. 
In fact, the diffusion coefficient is well approximated by 
\begin{eqnarray}
  D(\rho) &\sim &(\rho_{\rm g} - \rho)^\phi \,,
\end{eqnarray}                                
with $\phi \simeq 3.1$~\cite{KoAn}. 
The equilibrium with $\rho > \rho_{\rm g}$ is not reached by
 compaction after extremely long times:
if a chemical potential $\mu$ such that $\rho(\mu) > \rho_{\rm g}$
is applied, the system falls out of equilibrium.
Moreover, the density obtained at long times depends on the history: the 
slower the chemical potential is raised, the denser the system 
becomes~\cite{KuPeSe}.

We are here interested in this out of equilibrium dynamics: we
therefore perform a compression, starting from low density,
by raising the chemical potential up to a high value $\mu=3$.
Since the equilibrium density at $\mu=3$ 
is much larger than the jamming density $\rho_{\rm g}$, 
aging and very slow compaction ensue. 
We record the density $\rho(t)$, the density of
mobile particles $\rho_m(t)$, and the spatial  
structure function $g_{dyn}(r,t)$ defined as the probability that 
two sites at distance $r$ are occupied. 
Since we work at finite sizes, and since particles are always added
at the same layer, density heterogeneities do appear between the topmost
layer and the rest of the box if the compression is too fast. 
To avoid any systematic error, we use a slow compression ($\Delta t=10^4$ 
MC sweeps for each increase of $\Delta \mu=0.01$), and we measure the 
various quantities in the center of the box only, where we checked that 
the system is indeed homogeneous.

\begin{figure}[7]
\epsfxsize=3.4in
\centerline{\epsffile{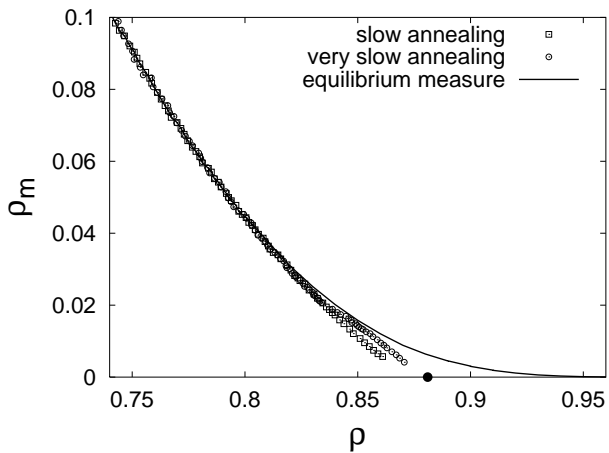}}
\vspace{0.2cm}
\caption{Parametric plot of the density of mobile particles versus
the density, for the equilibrium measure (data obtained by generating
at random configurations of a given density and measuring the number of
mobile particles), and during two compaction procedures (with
$\Delta \mu/\Delta t = 10^{-5}$ and $10^{-6}$).}
\label{fig:rhorhom}
\end{figure}

We show in Fig. \ref{fig:rhorhom} the parametric plot 
$\rho_m(t)$ versus $\rho(t)$; at short times and low density, 
it follows the equilibrium curve (obtained by generating at random
configurations of density $\rho$ and thus measuring the mean
density of mobile particles for the Gibbs measure); at low $\mu$ indeed,
the relaxation time of the system is smaller than the rate of increase
of $\mu$, so the system has time to equilibrate. 
As the density approaches $\rho_{\rm g}$
however, compaction slows down and $\rho_m(t)$ gets smaller. At
large times the system is approaching  
$(\rho \sim \rho_{\rm g},\rho_m \sim 0)$.

\begin{figure}[7]
\epsfxsize=3.4in 
\centerline{\epsffile{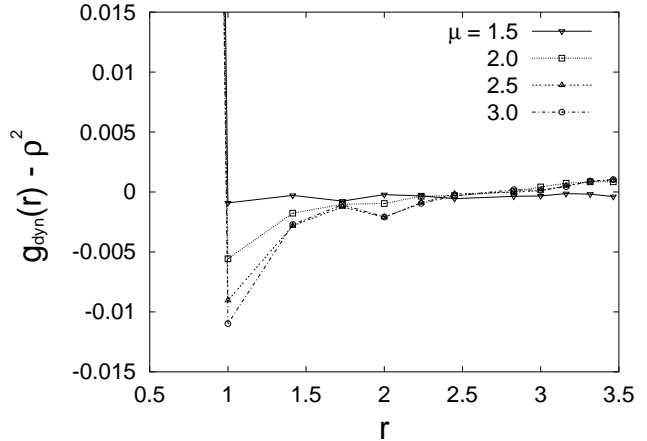}}
\vspace{0.2cm} 
\caption{Dynamic structure function obtained in a very slow
compression, at different times, i.e. different values of $\mu$; at 
$\mu=1.5$ the system is still at equilibrium and the correlation function
is therefore equal to $g_{equil}$, but stronger and stronger deviations
are observed as $\mu$ is raised.}
\label{fig:g_dyn_ka}
\end{figure}

We have also measured the dynamical structure function $g_{dyn}(r,t)$, 
displayed in Fig.
\ref{fig:g_dyn_ka};
induced and spontaneous displacements were measured and compared
in~\cite{Se}, and these data will be displayed and used in section III-F.

\subsection{The auxiliary model}

We introduce an `auxiliary model' which will allow us to define the
Edwards measure for the KA model.
In this model particles have energy equal to one if the dynamic rule
of the KA model allows them to move, and to zero otherwise. 
The Hamiltonian is therefore highly complicated, involving next-nearest 
neighbor interactions.
We can however introduce an auxiliary temperature $1/\beta_{aux}$ 
associated to the auxiliary energy $E_{aux}$ (equal to the number
of particles that are able to move) and 
perform a simulated annealing, at fixed number of particles: 
at low $\beta_{aux}$ all configurations are sampled uniformly, 
while, as $\beta_{aux}$ grows, the sampling is restricted to configurations 
with vanishing fraction of moving particles.
The Monte Carlo procedure uses non-local moves (accepted with a standard 
Metropolis probability 
$\min\left\{1,\exp(-\beta_{aux} \Delta E_{aux})\right\}$) 
which allow for an efficient sampling: these non-local moves have nothing 
to do with the true dynamics of the original model, and therefore the 
auxiliary model is not glassy..  
\begin{figure}[7] 
\epsfxsize=3.4in 
\centerline{\epsffile{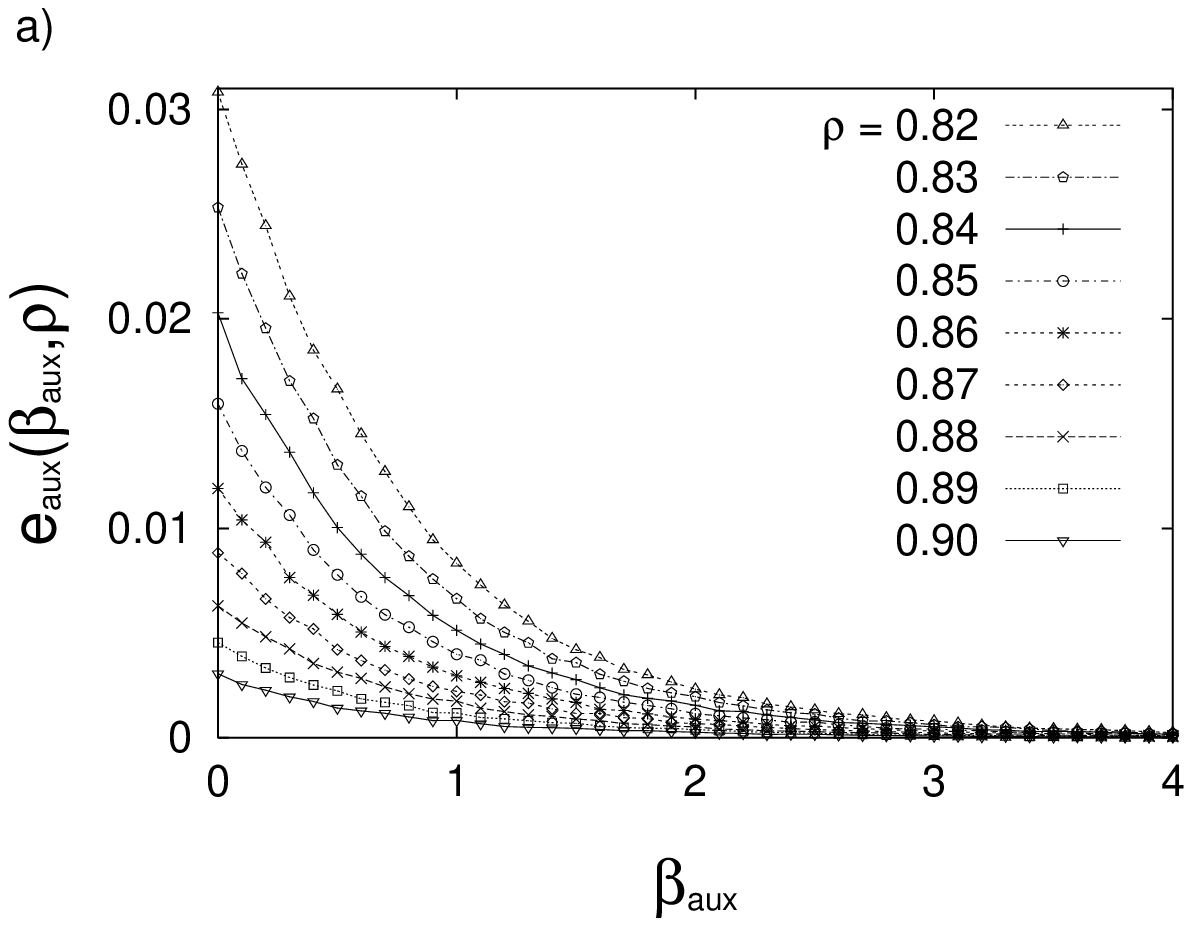}}
\vspace{0.75cm}
\epsfxsize=3.4in  
\centerline{\epsffile{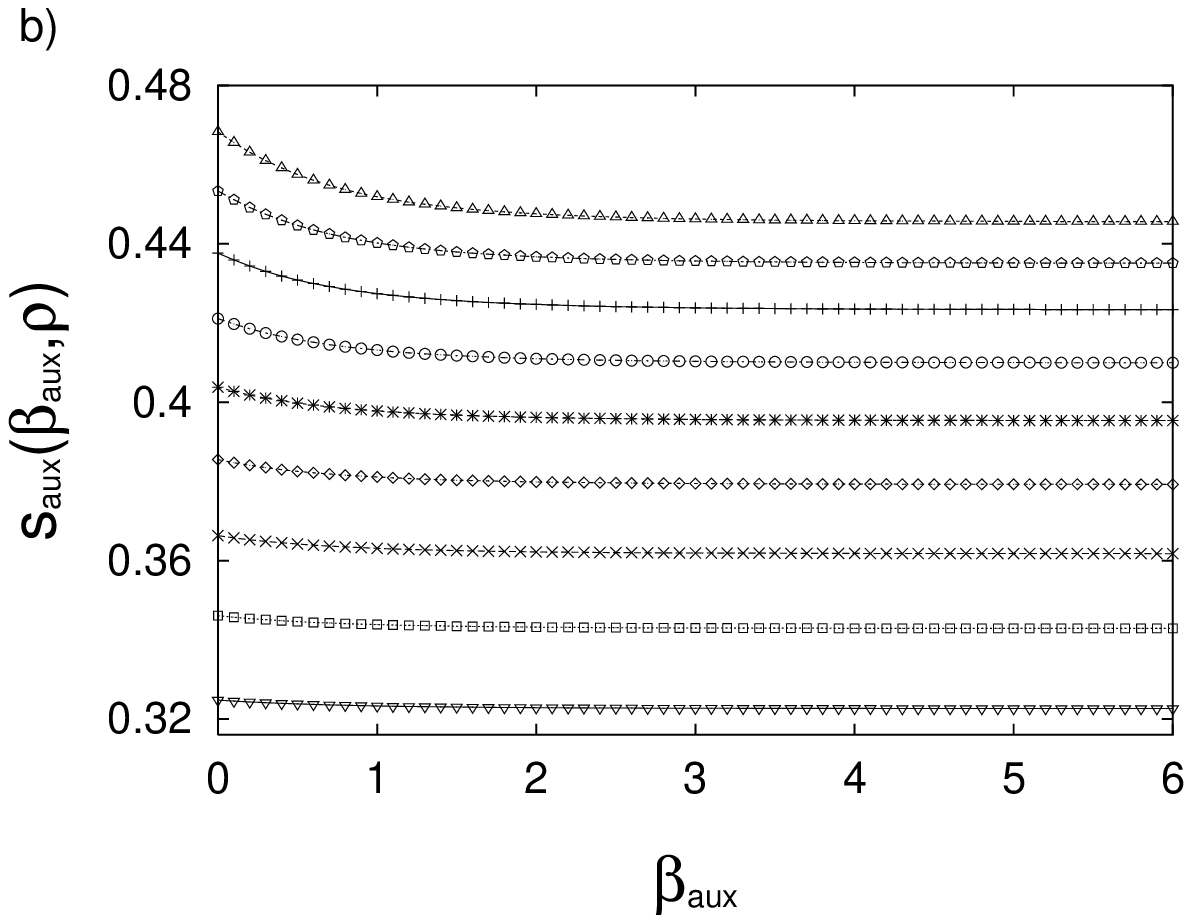}} 
\vspace{0.2cm} 
\caption{Thermodynamical properties of the auxiliary model.
a) Energy per particle $e_{aux}$  vs inverse 
temperature $\beta_{aux}$ at different particle density $\rho$.
b) Entropy per particle $s_{aux}$ as obtained from thermodynamic integration
of the energy data.
} 
\label{fig:ka_aux} 
\end{figure} 
In this way, we obtain the equilibrium energy density of the auxiliary model,
$e_{aux}(\beta_{aux},\rho)$ and its entropy density
$s_{aux}(\beta_{aux},\rho)$ by thermodynamic integration:
\begin{eqnarray}
  s_{aux}(\beta_{aux},\rho) &=& s_{equil}(\rho) +
  \beta_{aux} \, e_{aux}(\beta_{aux},\rho) \nonumber \\
  & &  
  - \int_0^{\beta_{aux}} e_{aux}(\beta_{aux}',\rho) \, d \beta_{aux}'   \,.
\end{eqnarray}
where we set
\begin{eqnarray}
  s_{aux}(0,\rho) &=& s_{equil}(\rho) 
\end{eqnarray}
since the limit $\beta_{aux} \to 0$ corresponds to the equilibrium 
measure.
In fig.~\ref{fig:ka_aux} we show a subset of data concerning the energy 
and entropy densities of the auxiliary model. 
(The energy has been computed in the range 
$\rho \in [0.65, 0.95]$ with a step in density $\Delta \rho =0.005$
and for $\beta_{aux} \in [0,20]$ with step $\Delta \beta_{aux} =0.1$). 

\begin{figure}[7]
\epsfxsize=3.4in 
\centerline{\epsffile{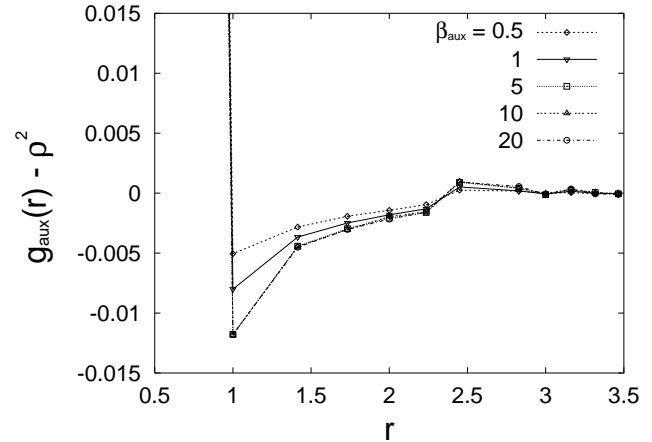}}
\vspace{0.2cm} 
\caption{Structure function obtained in the auxiliary model at different
values of the temperature $\beta_{aux}$ and for a density $\rho = 0.87$.
The data for $\beta_{aux}=5,\, 10$ and $20$ are indistinguishable.}
\label{fig:g_aux_ka}
\end{figure}

We also evaluate the structure function of the auxiliary model 
$g_{aux}(r,\beta_{aux})$ which is shown in fig.~\ref{fig:g_aux_ka}
for a density $\rho = 0.87$.
It is clear here as well as in fig.~\ref{fig:ka_aux}
that the limit $\beta_{aux} \to \infty$ considered in the next subsection
is already approached for $\beta_{aux} \simeq 5$.

\subsection{Edwards measure}

\begin{figure}[7]
\epsfxsize=3.4in
\centerline{\epsffile{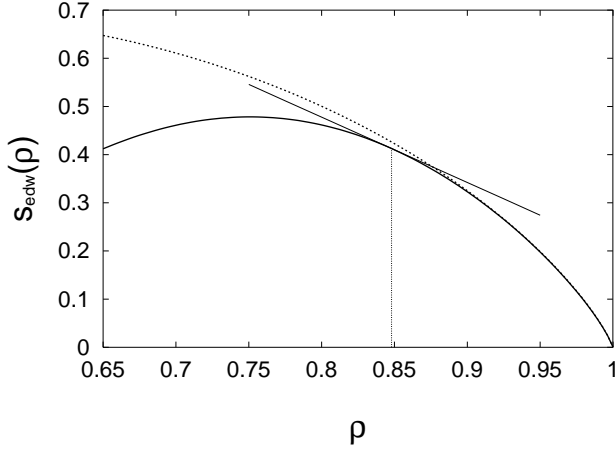}}
\vspace{0.2cm}
\caption{Edwards entropy per particle of the Kob-Andersen model vs. density
(full curve).
For comparison we also show the equilibrium entropy (dashed curve).
At high enough density  the curves are indistinguishable,
and join exactly only at $\rho=1$.
The slope of the tangent to $s_{edw}(\rho)$ for a generic $\rho$ allows to
extract $T_{edw}(\rho)$ from the relation
$T_{edw}(\rho) \frac{ds_{edw}}{d\rho}= \frac{ds_{equil}}{d\rho}$.
}
\label{fig:s_edw_ka}
\end{figure}

To evaluate the observables with Edwards' measure, i.e. the set of 
configurations where all particles are unable to move, we consider now 
the limit $\beta_{aux} \to \infty$ of the observables computed in the 
auxiliary model.
For example, the Edwards entropy is then obtained as
\begin{eqnarray}
  s_{edw}(\rho) & \equiv &
  \lim_{\beta_{aux} \to \infty} s(\beta_{aux},\rho)    \nonumber\\
  & = & s_{equil}(\rho)   
  - \int_0^{\infty} e_{aux}(\beta_{aux}',\rho) \, d \beta_{aux}'   \,,
\end{eqnarray}
since
\begin{eqnarray}
  \lim_{\beta_{aux} \to \infty} \beta_{aux} \, e_{aux}(\beta_{aux},\rho)
 & = & 0   \,.
\end{eqnarray}
In fig.~\ref{fig:s_edw_ka} we plot the Edwards and the equilibrium entropy
as a function of the particle density.

Comparison of Fig. \ref{fig:s_edw_ka} with Fig. \ref{fig:rhorhom} shows
that the most typical blocked configurations ($\rho \sim 0.75$)
are irrelevant as far as the compaction dynamics is concerned.

\begin{figure}[7]
\epsfxsize=3.4in 
\centerline{\epsffile{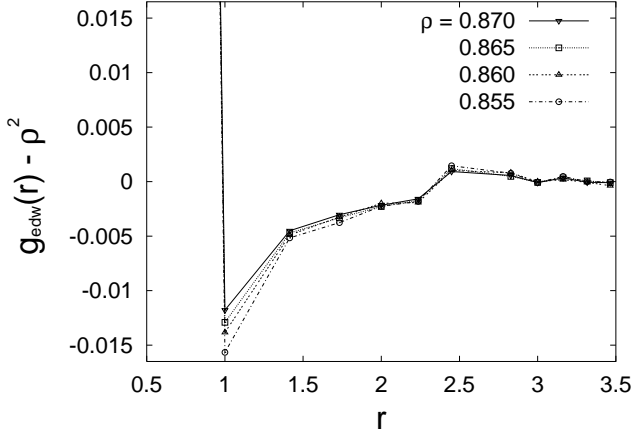}}
\vspace{0.2cm} 
\caption{Edwards structure function $g_{edw}(r)$ obtained as limit of the 
$g_{aux}(r,\beta_{aux})$ for $\beta_{aux} \to \infty$. As the density 
increases the deviation from $g_{equil}$ gets less pronounced.
}
\label{fig:g_edw_ka}
\end{figure}

Since the relation between chemical potential, temperature and entropy 
density at equilibrium is given by~(\ref{eq:mu_eq}),
the natural definition for Edwards' temperature is
\begin{eqnarray}
T_{edw}^{-1} &=& -\frac{1}{\mu} \, \frac{ds_{edw}(\rho)}{d\rho} \,.
\end{eqnarray}
However we work here at fixed density for Edwards' measure, and we therefore 
compute:
\begin{eqnarray}
  T_{edw} &=& 
  \frac{\frac{ds_{equil}(\rho)}{d\rho} }{ \frac{ds_{edw}(\rho)}{d\rho}}
\end{eqnarray}

Similarly, the Edwards measure structure function, $g_{edw}(r)$, 
is obtained as 
\begin{eqnarray}
  g_{edw}(r) & = & \lim_{\beta_{aux} \to \infty} g_{aux}(r,\beta_{aux}) \,.
\end{eqnarray}
and displayed in fig.\ref{fig:g_edw_ka} for various densities.

Two remarks are in order
\begin{itemize}

\item as the density approaches $1$, almost all particles become blocked
even for Gibbs' measure; more precisely, 
$\lim_{\rho \to 1} \rho_m^{Gibbs} = 0$; thus, Edwards' and Gibbs' measures
get closer, which is seen in fig.~\ref{fig:s_edw_ka}
 by the fact that the curves
for $s_{equil}$ and $s_{edw}$ get very close, 
and $\lim_{\rho \to 1} T_{edw}= T_{Gibbs} =1$. Similarly,
$g_{edw}(r)$ deviates less from $g_{equil}$ as $\rho$ increases,
as is shown in Fig.~\ref{fig:g_edw_ka}.

\item Edwards' measure is precisely
defined as a sampling over configurations with vanishing fraction
of mobile particles. We mention here for completeness
the straightforward generalization of Edwards' measures 
as the set of configurations
with fraction of mobile particles {\em smaller than} $\epsilon$
(cfr. the quasi-states of \cite{Frvi}); we then
can use the knowledge of $e_{aux}(\beta_{aux}, \rho)$ to define
$\beta^\epsilon$ as the value of the auxiliary temperature
such that $e_{aux}(\beta^\epsilon, \rho)=\epsilon$, and thus define 
\begin{eqnarray}
  s^\epsilon_{edw}(\rho) & \equiv &
  s(\beta_{aux}=\beta^\epsilon, \rho) \nonumber \\
  & = & s_{equil}(\rho) + \epsilon \, \beta^\epsilon \nonumber \\
  &  & - \int_0^{\beta^\epsilon} e_{aux}(\beta_{aux}, \rho )\,d\beta_{aux}\,.
\end{eqnarray}
It is clear that $s^1_{edw}=s_{equil}$, while $s^0_{edw}=s_{edw}$. 
We can also measure the structure factors 
$g^\epsilon(r)\equiv g(r,\beta^\epsilon)$. 
 
\end{itemize}

\subsection{Comparing the measures}

\begin{figure}[7] 
\epsfxsize=3.4in 
\centerline{\epsffile{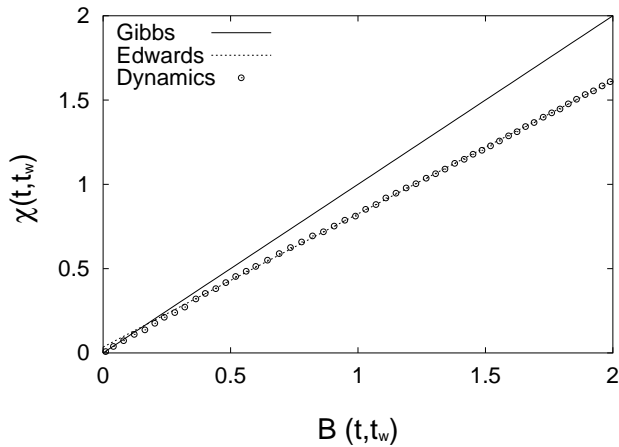}} 
\vspace{0.2cm} 
\caption{Einstein relation in the Kob-Andersen model: plot of  
the mobility   $\chi(t,t_w)$ vs. the mean-square displacement  
$B(t,t_w)$ (data shown as circles). 
The slope of the full straight line corresponds to the equilibrium temperature
 ($T=1$), and the slope of the dashed one to Edwards' prescription obtained 
from figure~\ref{fig:s_edw_ka} at $\rho(t_w)=0.848$. 
} 
\label{fig:x} 
\end{figure}

We are now in a position to compare the long-time results of the out of 
equilibrium dynamics with those obtained with the different
measures. Section II-C has already made clear that the equilibrium measure
is not able to describe these results.
Fig. \ref{fig:x} 
shows a plot of the mobility 

\begin{equation}
\chi(t,t_w)=\frac{1}{3N}\sum_{a=1}^{3} \sum_{k=1}^{N} \frac{\delta \left\langle (r_k^a(t) -
r_k^a(t_w)) \right\rangle }{\delta f} \ ,
\end{equation} 
obtained by the application of random forces to the particles
(see \cite{Se} for details),
{\em vs.} the mean square displacement

\begin{equation}
B(t,t_w)=\frac{1}{3N}\sum_{a=1}^{3} \sum_{k=1}^{N} \left\langle (r_k^a(t)-r_k^a(t_w))^2 
\right\rangle,
\end{equation}
testing in the compaction data the existence of a dynamical
temperature $T_{dyn}$~\cite{Se}. ($N$ is the number of particles 
and $a$ runs over the spatial dimensions)
One first remarks the existence of a dynamical 
temperature $T_{dyn}$. 
Furthermore the agreement between $T_{dyn}$ and $T_{edw}$,
obtained from the blocked configurations as in 
Fig.~\ref{fig:s_edw_ka}, for the density at which the dynamical
measurement were made, is clearly excellent.  
 
\begin{figure}[7] 
\epsfxsize=3.4in 
\centerline{\epsffile{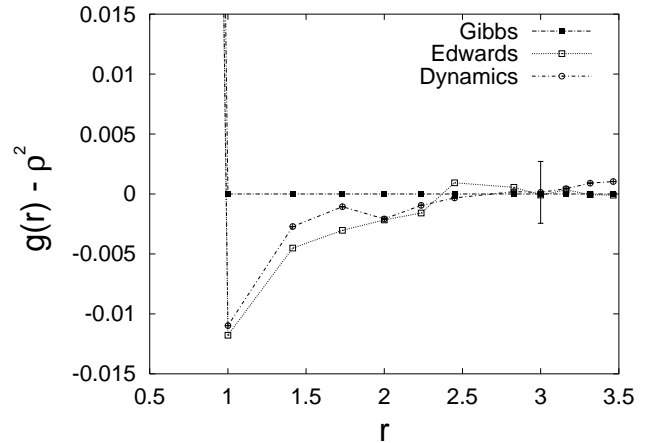}} 
\vspace{0.2cm} 
\caption{Structure functions $g(r) - \rho^2$ at 
density $\rho \simeq 0.87$ computed with the equilibrium, Edwards' and 
dynamical measure of the Kob-Andersen model. 
The three sets of data come from independent Monte-Carlo simulations. 
The dynamic structure function (circles) is obtained after a very slow 
compression by raising the chemical potential from $\mu=1$ to $\mu=3$ with 
an annealing rate of $10^{-6}$ Monte Carlo sweeps. 
The Edwards' structure function (open squares) is obtained from the 
auxiliary model. 
Although the equilibrium value of $g(r)-\rho^2$ is exactly $0$, we also 
obtain it by a Monte-Carlo simulation (full squares) in order to show that 
the difference in the short distance behaviour is not an artifact of the 
numerical simulation). The size of the typical error bar on dynamical data 
is shown at $r=3$. 
} 
\label{fig:g} 
\end{figure}

In Fig. \ref{fig:g} we plot the long-time dynamical $g_{dyn}(r,t)$,   
the equilibrium  $g_{equil}(r)=\rho^2$, and the Edwards' $g_{edw}(r)$ 
structure factors, for the same density
$\rho \sim 0.87$.  While $g_{equil}(r)$ is flat, the system has
developed during its dynamical evolution some structures,
which seem to be reproduced rather well by
$g_{edw}(r)$. Note that using the generalized Edwards' measures
does not improve the agreement in a clear-cut way because of the rather 
large error-bars on the dynamical data.

To summarize, during the compaction, the system falls out of equilibrium
at high density, and is therefore no more described by the equilibrium
measure. It turns out that Edwards' measure, constructed by a flat sampling of
the blocked configurations at the dynamically reached density, 
reproduces the physical quantities measured
at large times, and in particular predicts the correct value for the
dynamical temperature.

\section{Tetris Model}

In this section we extend the results obtained for the KA
model to another class of models, the so-called 
Tetris Model (TM)\cite{prltetris}.
We proceed as before by constructing explicitly 
Edwards' and Gibbs' measures, together with the corresponding
entropy and expectation values of some observables,
and comparing both sets of data with those obtained with
an irreversible compaction dynamics.
Notice that in this case the equilibrium measure is by no 
means trivial and it has to be computed numerically, using an auxiliary
model as for the construction of Edwards' measure.

\subsection{Model Definition}

The essential ingredient of the TM\cite{prltetris} 
is the geometrical frustration that for instance in granular packings 
is due to excluded volume effects arising from different shapes 
of the particles. This geometrical feature is captured in this 
class of lattice models where all the basic properties 
are brought by the particles and no assumptions are made on 
the environment (lattice). 
The interactions are not spatially quenched but are
determined in a self-consistent way by the local
arrangements of the particles. 
It is worth to notice how in this class of models
the origins of the randomness and of the frustration
coincide because both are given in terms of the particle
properties.

Despite the simplicity of their definition, these systems 
are able to reproduce many general features of granular media: 
the very slow density compaction\cite{prltetris}, segregation 
phenomena\cite{segtet}, dilatancy properties\cite{dilatancy} as well
as memory\cite{memory} and aging\cite{nicodemi,response}. 

\begin{figure}[h]
\centerline{
       \psfig{figure=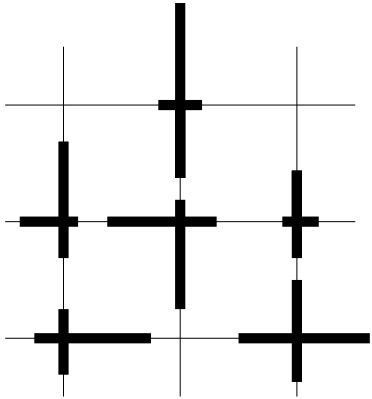,width=5cm,angle=0}
\vspace{0.5cm}}
\caption{Sketch of a local arrangement of particles in the 
Tetris Model: each particle can be schematized in general
as a cross with $4$ arms of different lengths,
chosen in a random way.}            
\label{partrtm}
\end{figure}

Let us recall briefly the definition of the model, which includes,
like in the real computer game {\em Tetris}, a rich variety 
of shapes and sizes. On a lattice each particle can be schematized 
in general as a cross with  $4$ arms (in general the number of arms is equals 
to the coordination number of the lattice) of different lengths, 
chosen in a random way. An example of particle configuration
on a square lattice is shown in Fig.~\ref{partrtm}.

The interactions among the particles obey the
general rule that one cannot have superpositions.
For instance one has to check that for two nearest-neighbor
particles the sum of the arms oriented along
the bond connecting the two particles is smaller
than the bond length. It turns out that in this way
the interactions between the particles are not fixed
once for all but they depend on the complexity of the
spatial configuration.

\begin{figure}[htb]
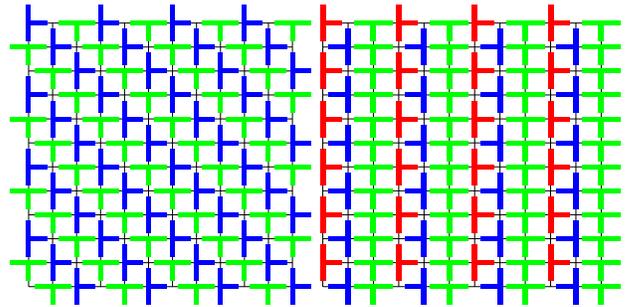

\centerline{
        \psfig{figure=rho2-3_12.epsi,width=4cm,height=4cm,angle=-90}
        \psfig{figure=rho2-3_12_bis.epsi,width=4cm,height=4cm,angle=-90}}
        \vspace*{0.5cm}
\caption{Two possible configurations of the Tetris model considered
with the maximal density $\rho= \rho_{max}= 2/3.$}  
\label{2_3conf}
\end{figure}

The extreme generality of the model definition allows a large 
variety of choices for the particles.
While the original model deals with simple rods, the fact that these rods
can arrange in an antiferromagnetic-like configuration
of density $1$ has motivated the use of random shapes to avoid
this pathology. In this
study, on the other side, we will use 
the so-called ``T''-shaped particles defined in such a way
that three arms have length equal to $\frac{3}{4} d$ and the 
fourth one zero length. $d$ sets the bond size on the square lattice. 
With this definition one has four types of particles corresponding to
the four possible orientation of the ``T''s on a square lattice.
Our choice has the following advantages: on the one hand,
no averaging over the disorder is needed; on the other hand, the process
of applying a chemical potential together for particles of various sizes
could produce a 'filtering' effect that would dynamically lead to
an artificially dense system with only small particles.
With the above given rules one can define the
allowed configurations. One can easily realize how the maximal 
allowed density  is equal to $\rho_{max} = \frac{2}{3}$
which corresponds to a number of possible configurations proportional
to the linear size of the lattice, $L$ (due to the
possible translations and symmetries).
Fig. \ref{2_3conf} shows two possible configurations with 
$\rho= \rho_{max}$.

\subsection{Equilibrium measure}

The equilibrium measure is obtained with an
annealing procedure. We can introduce a temperature 
$T=1/\beta$ associated with 
an energy $E$ defined as the total particle overlaps existing in a certain 
configuration. For each value of $T$ one 
allows the configurations with a probability given 
by $e^{-\beta E}$.
Starting with a large temperature $T$ (a very small $\beta$) 
one samples the allowed configurations by progressively 
decreasing $T$ (increasing $\beta$).
As $T$ is reduced $E$ decreases and only 
at $T=0$ (no violation of constraints allowed) 
the energy is precisely zero. The exploration of
the configuration space can be performed in two ways:

\begin{itemize}
\item[1)] working at constant density by interchanging the positions
of couples of particles. This procedure is used to compute 
$E(\beta,\rho)$ and
$e(\beta,\rho)$ (energy per particle), 
from which one can compute the 
equilibrium entropy per particle by the expression:

\begin{eqnarray}
s_{equil}(\rho) & \equiv & s_{equil}(\beta=\infty,\rho)=\nonumber\\
&= & s_{equil}(\beta=0,\rho) -  \int_0^\infty e(\beta,\rho) 
d\beta \,
\label{form_entro_equi}
\end{eqnarray}
For the choice made for the particles one has 
\begin{eqnarray}
s_{equil}(\beta=0,\rho) & = &
-\rho\ln\rho-(1-\rho)\ln(1-\rho)\nonumber \\
&& +\rho\ln 4,
\end{eqnarray}
which is easily obtained by counting the number of 
ways in which one can arrange
$\rho L^2$ particles of four different types on $N=L^2$ sites;

\item[2)] working at constant chemical potential $\mu$
by adding and removing particles. In this case the density fluctuates
and one measures $E(\beta,\mu)$. 
\end{itemize}

In both cases one can measure the particle-particle correlation
function (equivalent to the void-void correlation function)
$g_{equil}(r)$. Since, working at constant $\mu$ the density fluctuates,
we have always measured directly the ensemble average $<(g_{equil}(r)-\rho^2)>$
in order to make the data obtained with the two different methods
comparable. One can then compare the correlation function obtained 
at constant $\mu$ (which corresponds to a certain average density) with 
the correlation function obtained working at constant density.
The results of both methods are equivalent.

\begin{figure}[htb]
\centerline{
        \psfig{figure=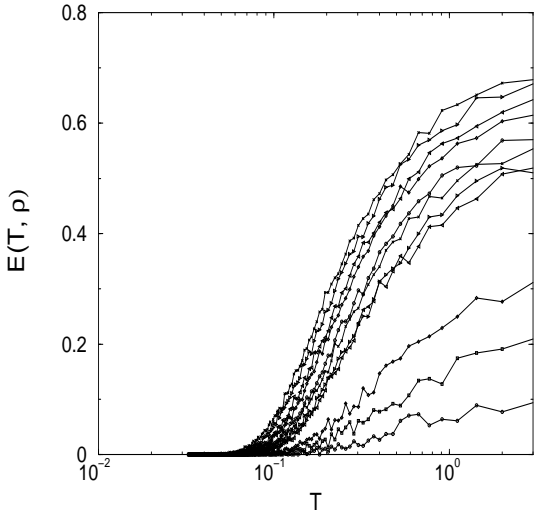,width=7cm,height=7cm,angle=-90}}
        \vspace*{0.5cm}
\caption{$e(\beta=1/T,\rho)$ for various values of 
$\rho$. From top to bottom the density decreases monotonically from 
$\rho=0.6$ to $\rho=0.1$}  
\label{energia_equi}
\end{figure}

\begin{figure}[htb]
\centerline{
        \psfig{figure=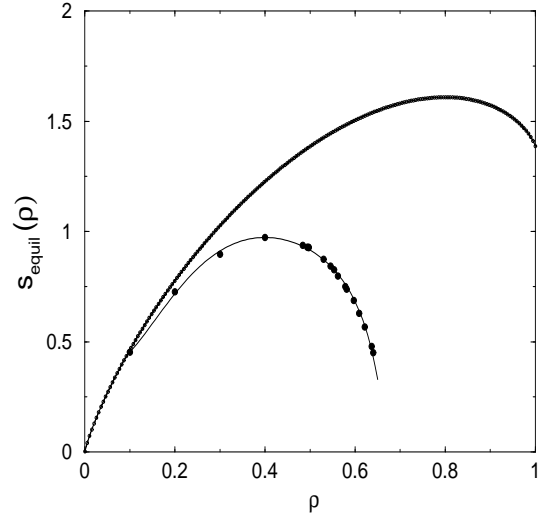,width=7cm,height=7cm,angle=-90}}
        \vspace*{0.5cm}
\caption{Entropy per particle in equilibrium $s_{equil}(\rho)$ and, 
for reference, entropy per particle at $T_{aux}=\infty$.
$s_{equil}(\rho)$ goes to zero at $\rho=\frac{2}{3}$.}
\label{entropia_equi}
\end{figure}

Fig.s (\ref{energia_equi}) and (\ref{entropia_equi}) 
report the results for $e(\beta,\rho)$ 
and $s_{equil}(\rho)$ as obtained from (\ref{form_entro_equi}).

\begin{figure}[htb]
\centerline{
        \psfig{figure=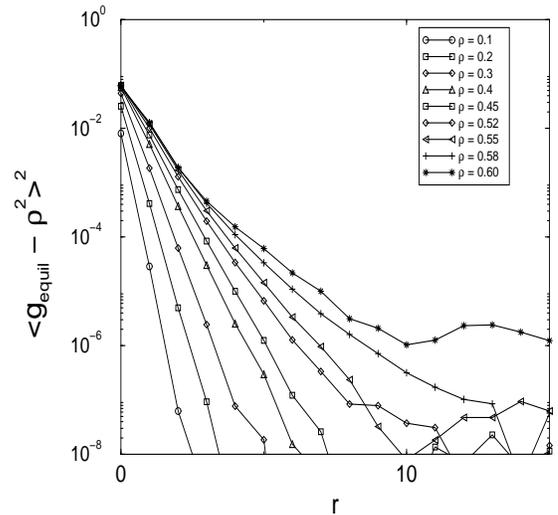,width=7cm,height=7cm,angle=-90}}
        \vspace*{0.5cm}
\caption{$<(g_{equil}(r)-\rho^2)>^2$ for various 
values of $\rho$. From top to bottom
the density decreases monotonically from 
$\rho=0.64$ to $\rho=0.1$.}  
\label{correl_equi}
\end{figure}

Fig. \ref{correl_equi}
reports the results for $<(g_{equil}(r)-\rho^2)>$
for different values of $\rho$. The correlation functions $g$ ($g_{equil}$ or 
$g_{edw}$ or $g_{dyn}$) actually
display oscillations around $\rho^2$, whose origin can be easily understood:
if a particle occupies a site, the exclusion rules decreases 
the probability that
a neighboring site will be occupied. We therefore plot in the
figures $<(g(r)-\rho^2)>^2$ in order to show the exponential decay
of the correlations.

\subsection{Edwards' measure}

Edwards' measure is obtained with an annealing procedure at fixed density.
This means that one samples the configurational space by interchanging 
the positions of couples of particles without violations of constraints.
In this way one is sampling the configurational space corresponding
already to $T=0$. In order to select only the subset of configurations
contributing to the Edwards' measure we introduce an auxiliary
temperature $T_{aux}$ (and the corresponding 
$\beta_{aux}=1/T_{aux}$)
and, associated to it, an auxiliary energy $E_{aux}$ which, 
for each configuration,
is equal to the number of mobile particles, in the same way as for the KA 
model. A particle is defined as mobile
if it can be moved according to the dynamic rules of the original model.

Let us describe in detail how the measurements are performed.
Since for each given density one is interested in the subset 
of the equilibrium configurations with a reduced particle mobility,
we start with an annealing procedure precisely identical to the one used
for the equilibrium measure. This procedure allows us to reach a starting
configuration with a given density and no constraints violated.
At this stage we perform a Monte Carlo procedure which exchanges the positions
of couples of particles without violation of constraints: 
this procedure accepts the non-local moves 
with a standard Metropolis probability 
$\min (1,\exp(-\beta_{aux} \Delta E_{aux})$) 
which allow for an efficient sampling. These non-local moves define an auxiliary
dynamics which has nothing to do with the true dynamics of the original model, 
and therefore the auxiliary model is not glassy.

As for the equilibrium we have performed a certain set of measures.

In particular we have measured $E_{aux}(\beta_{aux},\rho)$, i.e. the 
decrease of the auxiliary energy at fixed density. In order to do this
one performs an annealing procedure increasing progressively $\beta_{aux}$ 
and monitoring for each $\beta_{aux}$ the corresponding configurational
energy $E_{aux}(\beta_{aux},\rho)$. From this measure one can compute
the Edwards' entropy per particle defined by:

\begin{eqnarray}
s_{edw}(\rho) & \equiv & s_{aux}(\beta_{aux}=\infty,\rho)=\nonumber\\
 &= &s_{equil}(\rho) -  \int_0^\infty e_{aux}(\beta_{aux},\rho) 
d\beta_{aux} \,
\label{form_entro_edw}
\end{eqnarray}
where $e_{aux}(\beta_{aux},\rho)$ is the auxiliary Edwards' energy 
per particle and we put $s_{equil}(\rho)= s_{aux}(\beta_{aux}=0,\rho)$.

\begin{figure}[htb]
\centerline{
        \psfig{figure=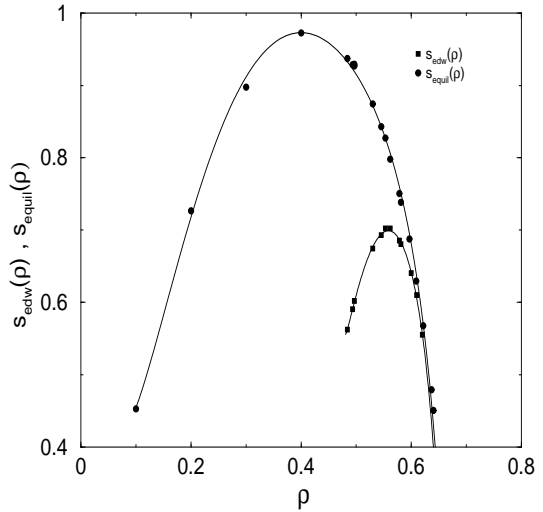,width=7cm,height=7cm,angle=-90}}
        \vspace*{0.5cm}
\caption{Edwards' entropy per particle, $s_{edw}(\rho)$  
and equilibrium entropy per particle, $s_{equil}(\rho)$. Both
go to zero at $\rho=\frac{2}{3}$.
}  
\label{entropia_edw}
\end{figure}

Fig. \ref{entropia_edw} 
reports the results for  $s_{edw}(\rho)$ as obtained from  
(\ref{form_entro_edw}) compared with $s_{equil}(\rho)$.

\begin{figure}[htb]
\centerline{
        \psfig{figure=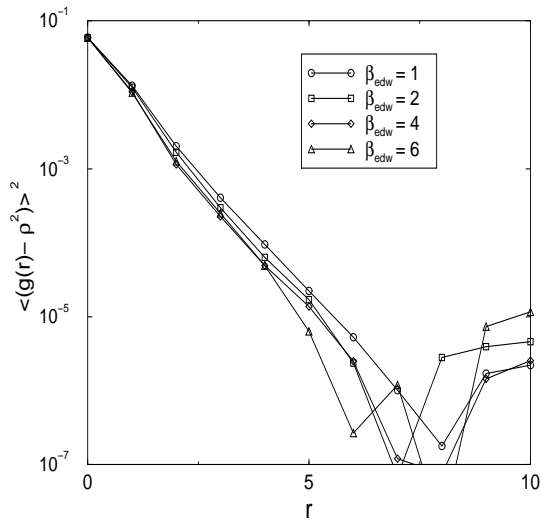,width=7cm,height=7cm,angle=-90}}
        \vspace*{0.5cm}
\caption{$<(g_{edw}(r)-\rho^2)>^2$ for $\rho=0.58$;  
$\beta_{aux}=1,2,4,6$. The behaviour seems to saturate
to a limit that should correspond to $\beta_{aux} \rightarrow \infty$ 
already for $\beta_{aux}$ of the order of $6$.}  
\label{corr_edw}
\end{figure}

For the computation of the particle-particle correlation function
we have to use a different strategy. Always starting from a configuration 
with a given density and no constraints violated one performs a
Monte Carlo procedure, at $\beta_{aux}$ fixed, which exchanges the positions
of pairs of particles without violation of constraints.
Each single simulation uses $\rho$ and $\beta_{aux}$ as input parameters.
In practice one is trying to sample all the configuration of density 
$\rho$ and with a particle mobility defined by  $\beta_{aux}$.
In this context the Edwards' prescription  should correspond to the limit
$\beta_{aux} \rightarrow \infty$.
In this way we have computed $<(g_{edw}(r)-\rho^2)>$ for several values
of $\beta_{aux}$ and we report the results in Fig. \ref{corr_edw}.
As one can see the limit $\beta_{aux} \rightarrow \infty$ is attained 
already for $\beta_{aux}$ of the order of $6$.

\subsection{Irreversible Compaction Dynamics}

\begin{figure}[htb]
\centerline{
        \psfig{figure=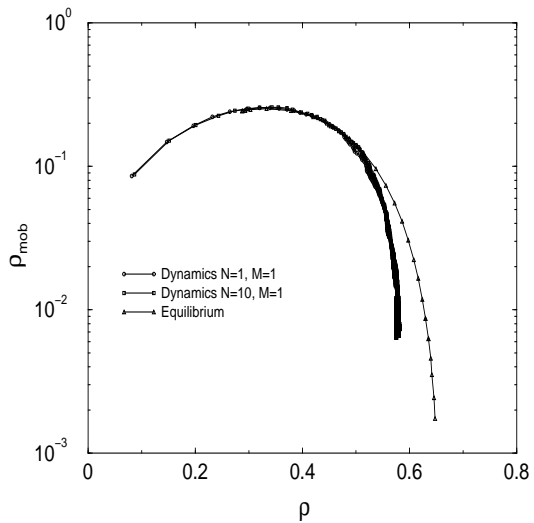,width=7cm,height=7cm,angle=-90}}
        \vspace*{0.5cm}
\caption{$\rho_{mob}$ vs. $\rho$ for two irreversible 
dynamics with $N=1,M=1$ and $N=10,M=1$.
For reference is plotted the equilibrium curve.}  
\label{rhomob_vs_rho}
\end{figure}

We now turn to the out of equilibrium dynamics of compaction, simulated by
starting from an empty lattice and performing
$N$ steps of attempted particle additions followed by 
$M$ steps of attempted particle diffusions. 
Fig. \ref{rhomob_vs_rho} reports the results for the 
density increase and for the fraction of mobiles particles as a function
of the density for $N=1,M=1$ and $N=10,M=1$. While
the limit $N=$ fixed $M -> \infty$ should coincide with the equilibrium,
it is clear that the system, at finite $M$, is not able to approach
its maximum density and falls out of equilibrium.
Here again the density at which the number of blocked configurations
is maximal is smaller than the one achieved by compaction.

It is particularly interesting to notice that in the out-of-equilibrium
configurations visited during the irreversible dynamics the fraction of mobile particles
$\rho_{mob}$ at fixed density is systematically smaller than 
the corresponding value 
in equilibrium. This suggests the possibility of distinguishing
between equilibrium and 
out-of-equilibrium configurations by looking at the spatial 
organization of the particles in both cases. 
We have then measured, during the compaction dynamics, the particle-particle
correlation function at fixed density. 
In the next section we shall
compare these results with those obtained on the basis of the 
equilibrium and Edwards' measures.


\subsection{Comparing different measures}

We are now able, as for the KA model, to compare Gibbs' and Edwards' measures
with the results of the out-of--equilibrium dynamics at large times.

\begin{figure}[htb]
\centerline{
        \psfig{figure=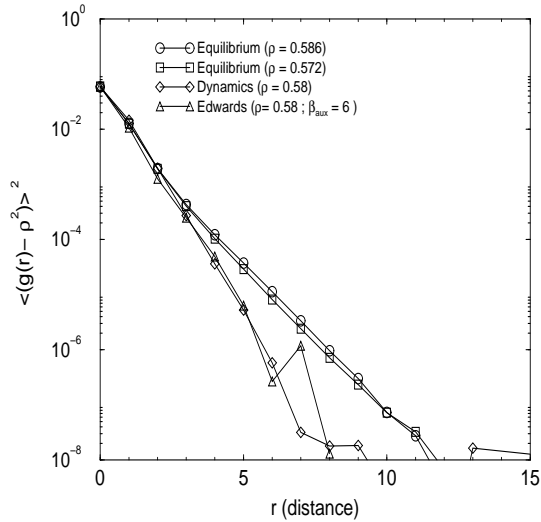,width=7cm,height=7cm,angle=-90}}
        \vspace*{0.5cm}
\caption{Comparison between the correlation functions obtained 
with the equilibrium measure, the Edwards measure ($\beta_{aux}=6$) 
and the irreversible dynamics ($N=1,M=1$). In all cases the system 
is considered at a density of $\rho \simeq 0.58$.}  
\label{correl_confr}
\end{figure}

In Fig. \ref{correl_confr} we plot the deviations of the 
particle-particle correlation functions from the uncorrelated value
$\rho^2$. In particular we compare $<(g_{dyn}(r)-\rho^2)>^2$ 
obtained during 
the irreversible compaction ($N=1, M=1$) with the corresponding 
functions obtained with the equilibrium and Edwards' measures.
It is evident that the correlation function, as measured
during the irreversible compaction dynamics, is significantly different from 
the one obtained with the equilibrium measure. On the other hand the correlation
functions obtained with Edwards' measure are able to better describe 
what happens during the irreversible dynamics. In particular 
what is observed is that
the correlation length seems to be smaller for configurations 
explored by the irreversible dynamics than in
the equilibrium configurations.
This aspect is captured by Edwards' measure which selects, 
the better the larger 
$\beta_{aux}$, configurations with a reduced particle mobility.
In practice one can summarize the problem as follows: given a certain density,
one can arrange the particles in different ways. The different configurations
obtained in this way differ in the particle mobility and this feature
is reflected by the change in the particle-particle correlation properties.

Also in this case, as in the KA example, it turns out that Edwards' measure, 
constructed by a flat sampling of the blocked configurations, is able to 
reproduce the physical quantities measured at large times. Investigations
are currently running to check also in the Tetris model whether the 
temperature predicted with the Edwards' approach coincide with 
the dynamical temperature $T_{dyn}$ as defined for the KA model.

\section{The 3-dimensional Random Field Ising Model at low temperature}

In this section, we consider a case in which Edwards' ensemble does
not give good results: the low temperature domain growth dynamics  
of a 3D Ising model in a weak random magnetic field. This model has
been applied in many different contexts
\cite{Nattermann}, and in particular in relation
with glasses \cite{alberici}.
The logarithmic relaxations it displays at low temperature, as well
as the dependence on its thermal history (like the influence of
the cooling rate)
can also induce comparisons with granular compaction.

The model is defined as usual: $N=L^3$
Ising spins ($s_i=\pm 1$) sitting on the sites of
a regular lattice of linear size $L$,
interact ferromagnetically, in
a random external field. The Hamiltonian is
\begin{equation}
H = - J \sum_{<i,j>} s_i s_j - \sum_i h_i s_i \ ,
\label{eq:hrfim}
\end{equation}
where the sum runs over pairs of nearest neighbours.

The strength of the ferromagnetic interaction $J$ can be set to $1$,
and the distribution of the random fields $h_i$ will be taken as
bimodal $h_i = \pm h_0$.
At high temperature, the system is in a paramagnetic phase; at low 
temperature and weak magnetic field, 
there exists a ferromagnetic phase (see \cite{Nattermann}).
The typical equilibrium configurations are therefore magnetized.

In the absence of random fields, the low temperature dynamics is
the well known coarsening of domains of plus or minus spins, whose typical
size grows as a power of time. 
In the case of the RFIM, the domain walls are pinned by the field, and the  
dynamics proceeds by thermal activation. The size of the domains therefore
grows only logarithmically with time. Moreover, the fact that
thermal barriers are easier to overcome at not too low temperatures
induces a strong dependence on the cooling rate. 
As above, we are interested in the limit of low but non-zero
temperatures.

In a large system, the long-time configurations obtained dynamically
are intertwined domains
of `up' and `down' spins having similar volumes, 
the global magnetization being zero. 
This is quite different from the equilibrium configurations 
at the same energy, which are magnetized. 
In fact, an easy way to show that the long-time dynamical configurations
differ from the equilibrium ones is to look at the distribution
of the magnetizations $P(M)$ in both cases: at equilibrium $P_{equil}(M)$ is peaked
around $\pm M(T)$, with $M(T) > 0$, while for the dynamics one obtains
at any finite time for $P_{dyn}(M)$
a single peak around $M=0$: in this domain growth dynamics, the system
does not choose at any finite time between the two basins of
attractions of the two ground states \cite{kurchan-laloux}.

The dynamics proceed by thermal activation; therefore the
long-time configurations are 'blocked', in the sense that, at zero
temperature, the system would be unable to escape from them. 
The question in the present context is now whether these 
'blocked configurations' are typical, i.e. if their characteristics
are well reproduced by a flat sampling of {\em all}
blocked configurations of the same energy.
We have therefore considered
the corresponding auxiliary model, in order to obtain this flat sampling.

\begin{figure}[htb]
\centerline{
        \psfig{figure=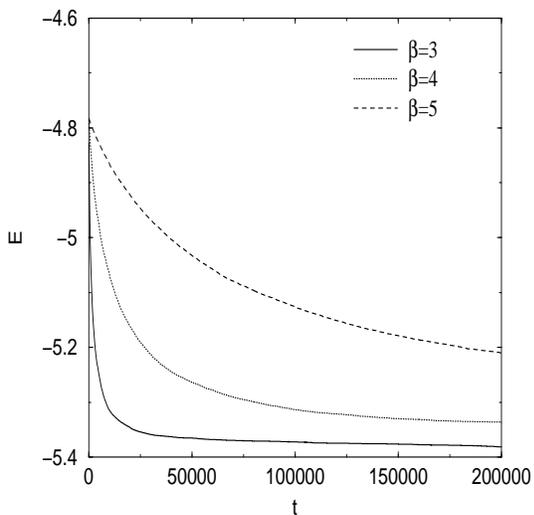,width=7cm,height=7cm,angle=-90}}
        \vspace*{0.5cm}
\caption{Energy versus time for the dynamics of the RFIM at inverse
temperatures $3$, $4$ and $5$. The energy decreases slower for larger
$\beta$ because the dynamics is activated. For one given sample
$E(t)$ would be a succession of plateaus, here the curves correspond to
an average over $64$ realizations of the random field.}
\label{e_rfim}
\end{figure}

Since we want to study the configurations at a given energy,
the auxiliary model has two terms: the first one
is quadratic, constraining the energy (\ref{eq:hrfim}) around a given
$E_0$, and the second one is the 
number of spins  not aligned with their local field, i.e. the number of
spins that can flip without thermal activation:
\begin{eqnarray}
E_{aux} & = & \beta_1 \left(- J \sum_{<i,j>} s_i s_j - \sum_i h_i s_i - E_0
\right)^2 +  \nonumber\\
&& \beta_2  \sum_i \Theta(-s_i {\cal H}_i) \ \ ,
\end{eqnarray}
where $\Theta$ is the Heavyside function, and
${\cal H}_i= J \sum_{<j,i>} s_j + h_i$
is the local field at site $i$. The two auxiliary inverse temperatures
$\beta_1$ and $\beta_2$ are used to perform an annealing, starting from
a random initial configuration of high energy. We use a single-spin flip
Metropolis algorithm, accepting the moves with probability
$\min (1,\exp(-\Delta E_{aux}))$. 
$E_0$ is taken negative and slightly larger
than the ground state mean energy $E_{GS}=-6 J N$, in order to
look at configurations having the same energy than the long time
dynamics configurations (the evolution of the energy during the
dynamical evolution is displayed in Fig. \ref{e_rfim}).

\begin{figure}[htb]
\centerline{
        \psfig{figure=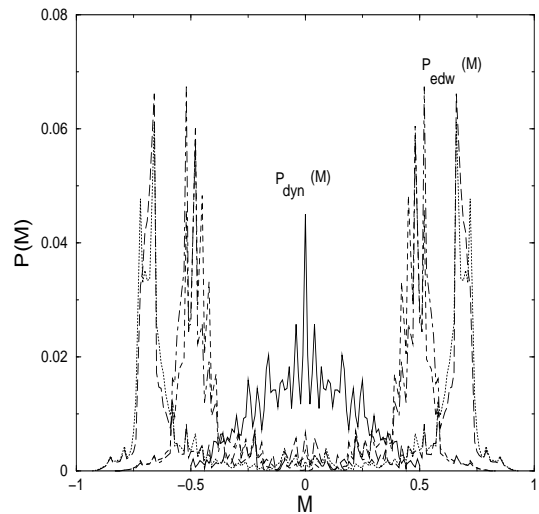,width=7cm,height=7cm,angle=-90}}
        \vspace*{0.5cm}
\caption{Histograms of the magnetization of the visited configurations
during the low temperature dynamics (full line: $P_{dyn}(M)$ for
$L=20$, $\beta=4$, i.e.
the energy per site of the system is between $-5.2$ and $-5.4$; the finite
width of the peak comes from the relatively small size of the sample) and
for Edwards' measure at $E_0=-5.2$ and $-5.4$ (dotted, dashed, long-dashed
and dot-dashed lines: $P_{edw}(M)$
for $L=20$ and $\beta_2=4$ and $6$; $E_0=-5.4$
corresponds to the peaks at higher $M$). The dynamics samples
configurations with low magnetization, while the typical blocked
configurations are magnetized.}
\label{pm_rfim}
\end{figure}

The first, simplest observable to look at
is the distribution of the magnetization, $P_{edw}(M)$, averaged
over realizations of disorder. 
Typical results are displayed in Fig. \ref{pm_rfim}, for
the $P(M)$ obtained dynamically or with a sampling of the
blocked configurations, for a system of $20^3$ spins, and
for $E_0 = -5.4 J N$ and $-5.2 J N$ (we have also simulated
systems of $30^3$ spins). While the dynamical $P_{dyn}(M)$
consists in a single peak around $M=0$ (getting narrower for larger
system sizes), the distribution
$P_{edw}(M)$ is clearly bimodal, presenting two peaks around two
symmetric values of the magnetization. The precise values of the peaks
depend on $E_0$ (decreasing for increasing $E_0$:
it is clear that, for too high values of $E_0$, the 
configurations are less magnetized), and their width depend
on system size, getting narrower for larger systems.

It appears therefore that the configurations dominating Edwards' 
distribution are {\em magnetized}.
(Note that a similar result was obtained in
\cite{dean} for the low-energy metastable states
for a ferromagnet on random thin graphs).
 At variance with the previously
studied models (KA and Tetris), Edwards' distribution is therefore
unable to describe the typical configurations obtained dynamically. 

The example of the RFIM clearly underlines the difference with the inherent
structure construction of Kob et al. On the one hand, we have seen that
Edwards' measure does not reproduce the dynamic results. On the other hand,
it is clear that different thermal histories, just as those considered in
\cite{KoScTa}, would yield domain configurations with essentially the
same shapes if the end energy is the same, and zero magnetization,
contrary to the typical configurations at that energy.

\section{`Chaoticity' properties}

We have shown two models for which Edwards' construction gives a good
approximation, and one for which it does not.
What is the distinguishing feature between them?
 A distinction one can make, suggested by glass 
theory~\cite{baldassarri,cude,bamebu,Frvi}, is obtained by 
studying their 
`chaoticity' properties as follows: after aging for a time $t_w$,  
two copies (clones) are made of the system, 
and allowed to evolve subsequently 
with different realizations of the randomness  in the updating
 procedure.
The question is then whether the trajectories diverge or not.
Note that for this criterion to make sense, it should always be applied 
at non-zero (though weak) tapping or shearing. 
 
 The results summarized below seem to indicate that the
 condition of chaoticity is necessary. It is  however not sufficient:
 Bouchaud's
 `trap model' \cite{review} is chaotic but its fluctuation-dissipation
 properties are not directly related to the density of states \cite{Bou}.

For the three models discussed in this paper, 
we have measured the normalized average overlap $Q_{t_w}(t)$ defined as
\begin{itemize}
\item for the KA model
\begin{equation}
Q_{t_w}(t) = \langle \frac{1}{N} \sum_i n_i(t) n_i^{clone}(t) 
- \rho(t) \cdot \rho_{clone}(t) \rangle
\end{equation}
where $n_i$ (resp. $n_i^{clone}$)
is $1$ if there is a particle on the site $i$ for
the original model (resp. for the clone), and $0$ if the site is empty;

\item for the Tetris model
\begin{equation}
Q_{t_w}(t) = \langle \frac{(\frac{1}{N}
\sum_{i,j} A_{i,j}(t) - \frac {\rho(t) \cdot \rho_{clone}(t)}{4})}
{(\rho_{clone}(t) - \frac{\rho_{clone}(t)^2}{4})} \rangle
\label{chaos}
\end{equation}
where $A_{i,j}(t)$ is a function that gives one if on the site $(i,j)$ the 
two copies present the same particles and zero otherwise. 
We have four types of particles in the system and this is the reason
for the factors $4$ in (\ref{chaos}).

\item for the RFIM
\begin{equation}
Q_{t_w}(t) = \langle \frac{1}{N} \sum_i s_i(t) s_i^{clone}(t) \rangle
\end{equation}

\end{itemize}

 The brackets indicate
an average over different realizations of the randomness.

\begin{figure}[htb]
\centerline{
        \psfig{figure=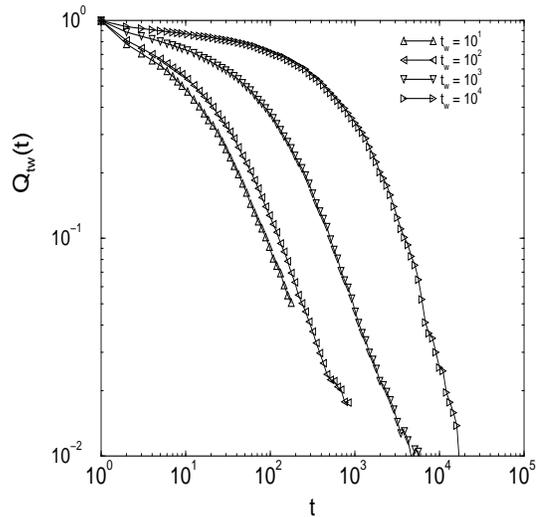,width=7cm,height=7cm,angle=-90}}
        \vspace*{0.5cm}
\caption{Mean overlap $Q_{t_w}(t)$ 
between two clones in the Tetris model: the two clones are separated at $t_w$
and evolve subsequently with different noises. $Q_{t_w}(t)$ always decreases
to zero (the slower the larger $t_w$), showing that the clones always 
diverge.}  
\label{chaos_fig}
\end{figure}

\begin{figure}[7] 
\epsfxsize=3.4in 
\centerline{\epsffile{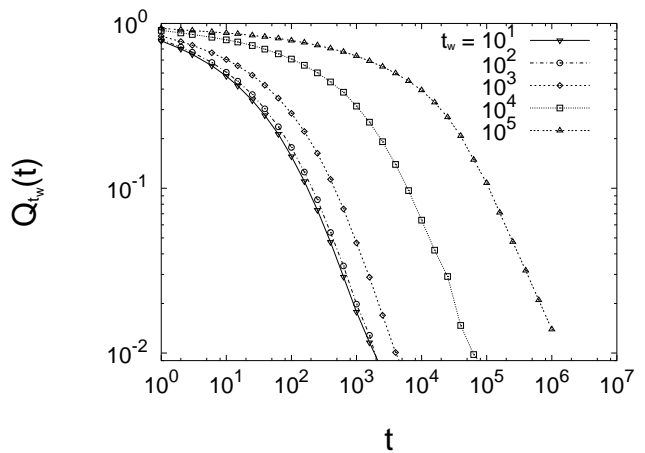}} 
\vspace{0.2cm} 
\caption{Same as Fig. {\protect \ref{chaos_fig}} for the KA model.
} 
\label{fig:clone_ka} 
\end{figure}

Fig. \ref{chaos_fig} and \ref{fig:clone_ka}
report the results for $Q_{t_w}(t)$
for several values of $t_w$, respectively for the Tetris and the KA model.
It is clear that in both cases
the two copies of the system always tend to diverge. This is at variance
with what happens in the domain growth dynamics \cite{bamebu}, and in 
particular for the RFIM, where we have checked that 
the two copies retain a finite overlap at all the times studied.

\section{Conclusions}

In summary, we have proposed a simple and systematic procedure to
construct a flat sampling of the `blocked configurations',  i.e. to
calculate averages with
 Edwards' measure. We have shown, for two representative 
finite-dimensional models, that this measure gives different results
than the equilibrium measure, and is able to
reproduce the dynamical sampling of the out-of-equilibrium
compaction dynamics for various observables.
The connection of Edwards' ensemble with the dynamical FDT temperature
immediately suggests experiments to check the validity of
these ideas, for example by studying diffusion and mobility 
of different tracer particles within driven granular media. 

At present, the 
correspondence between Edwards' distribution and the long-time dynamics 
is at best checked but does not follow from any known principle. 
Now that several concrete examples have lent credibility to Edwards'
construction, an effort to understand why it does in some
cases work and what is its range of validity has become  worthwhile.

 There remains the question of    generalizing
Edwards' measures in two directions: by considering  a fraction
$\epsilon>0$ of mobile particles, 
and by conditioning the   flat measure  to 
more macroscopic observables in addition to  density and energy.

V.L. and M.S. acknowledge 
the hospitality of the ESPCI of Paris where part of this work was 
carried out. MS is supported by a Marie Curie fellowship of 
the European Commission (contract ERBFMBICT983561).
This work has also been partially supported from the 
European Network-Fractals under contract No. FMRXCT980183.


\begin{thebibliography}{99}
 
\vspace*{-0.3in}
 
\begin{small}
\bibitem[*]{umr}Unit{\'e} Mixte de Recherche UMR 8627.
\end{small}
 
\vspace*{0.1in}

\bibitem{Nagel}
  H.M. Jaeger, S.R. Nagel,
  {\em Science} {\bf 255}, 1523 (1992).

\bibitem{Nagel2}
H.M. Jaeger, S.R. Nagel, and R.P. Behringer,
{\em Rev. Mod. Phys} {\bf 68}, 1259 (1996). 

\bibitem{Knight}
J. B. Knight, C. G. Fandrich, C. N. Lau, H. M. Jaeger, 
and S. R. Nagel, {\em Phys. Rev. E} {\bf 51}, 3957 (1995).

\bibitem{Sam}   
  S.F. Edwards, 
  in: {\it Granular Matter: An Interdisciplinary Approach},   
  A. Mehta, Ed. (Springer-Verlag, New York, 1994), and references therein.   

\bibitem{anita} 
  A. Mehta, R.J. Needs, S. Dattagupta,  
  {\it J. Stat. Phys.} {\bf 68}, 1131 (1992). 

\bibitem{Repo} 
  R. Monasson, O. Pouliquen,  
  {\it Physica A} {\bf 236}, 395 (1997). 

\bibitem{bklm} A. Barrat, J. Kurchan, V. Loreto and M. Sellitto,
{\em  Phys. Rev. Lett.} {\bf 85} 5034 (2000).

\bibitem{Struik} 
  See, for example Chapter 7 of: L.C.E.  Struik,  
  {\it Physical Ageing in Amorphous Polymers and Other materials}, 
  (Elsevier, Houston, 1978). 

\bibitem{KoAn}
  W. Kob, H.C. Andersen,
  {\it Phys. Rev. E} {\bf 48}, 4364 (1993). 

\bibitem{prltetris}
E. Caglioti, V. Loreto, H.J. Herrmann and M. Nicodemi,
{\em Phys. Rev. Lett.} {\bf 79}, 1575 (1997).

\bibitem{Nattermann} T. Nattermann,
  in {\it Spin-glasses and random fields}, A. P. Young, Ed.
  (World Scientific, Singapore, 1997).

\bibitem{Ott} 
E. Ott, {\it Chaos in Dynamical Systems} 
(Cambridge University Press, Cambridge 1993).


\bibitem{KTW} 
  T.R. Kirkpatrick, D. Thirumalai,  
  {\it Phys. Rev. B} {\bf 36}, 5388 (1987). 
 
\bibitem{KTW2}  
  T.R. Kirkpatrick, P. Wolynes,  
  {\it Phys. Rev. A} {\bf 35}, 3072 (1987). 
 
\bibitem{review}  
  J.-P. Bouchaud,  L.F. Cugliandolo, J. Kurchan and M. M\'ezard, 
  in {\it Spin-glasses and random fields}, A. P. Young, Ed. 
  (World Scientific, Singapore, 1997). 
 
\bibitem{Cukupe}  
  L.F. Cugliandolo, J. Kurchan, L. Peliti, 
 {\it Phys. Rev. E} {\bf 55}, 3898 (1997). 
 
\bibitem{remi}  
  R.~Monasson,  
  {\it Phys. Rev. Lett.} {\bf 75}, 2847 (1995). 
 

\bibitem{jamming} 
  J. Kurchan,  
  in {\it Jamming and Rheology: Constrained Dynamics 
  on Microscopic and Macroscopic Scales} (1997), 
  http://www.itp.ucsb.edu/online/jamming2/, 
  and Edwards, S.F., Liu, A. and Nagel, S.R. Eds.,   
  to be published, cond-mat 9812347. 
 
\bibitem{Theo}  
  Th.M. Nieuwenhuizen,  
  {\it Phys. Rev. E} {\bf 61}, 267 (2000). 

\bibitem{Frvi}  S. Franz, M.A. Virasoro, 
  {\it J. Phys. A} {\bf 33}, 891 (2000). 
 
\bibitem{Felix}  
  A. Crisanti,  F. Ritort, cond-mat/9911226. 
 
\bibitem{biroli} G. Biroli and J. Kurchan, {\em Phys. Rev. }{\bf E},
  to be published.

\bibitem{jorge-trieste}  
  J. Kurchan, {\it J. Phys. Condensed Matter},  
  {\bf 12}, 6611 (2000). 
 
\bibitem{KoScTa}
  W. Kob, F. Sciortino, P. Tartaglia,
  {\em Europhys. Lett.} {\bf 49}, 590 (2000). 


\bibitem{inherent_equilibrium}
  P. H. Stillinger and T. A. Weber, 
{\em Phys. Rev.} {\bf A25}, 978 (1982); S. Sastry, P. G. Debenedetti
 and F. H. Stillinger, {\em Nature} {\bf 393}, 554  (1998);
  B. Coluzzi, G. Parisi and P. Verrocchio,
  {\em Phys. Rev. Lett.}{\bf 84}, 306 (2000);
  F. Sciortino, W. Kob, and P. Tartaglia, {\em Phys. Rev. Lett.}{\bf
    83}
3214 (1999).

\bibitem{CoNi} A. Coniglio and M. Nicodemi, cond-mat/0010191.

\bibitem{gotze}
For a review, see:
W. G\"otze, in  `Liquids, freezing and
glass transition',  J.P. Hansen, 
D. Levesque, J. Zinn-Justin Editors,
Les Houches (1989) (North Holland).

\bibitem{KuPeSe}
  J. Kurchan, L. Peliti, M. Sellitto,
  {\it Europhys. Lett.} {\bf 39}, 365 (1997).

\bibitem{Se}
  M. Sellitto,
  {\it Euro. J. Phys. B} {\bf 4}, 135 (1998).

\bibitem{SeAr}
  M. Sellitto, J.J. Arenzon,
  {\it Free-volume kinetic models of granular matter},
  {\it Phys. Rev. E} (December 2000).
 
\bibitem{response} A. Barrat, V. Loreto, 
{\em J. Phys. A.} {\bf 33}, 4401 (2000)

\bibitem{segtet}
E. Caglioti, A. Coniglio, H.J. Herrmann, V. Loreto and M. Nicodemi,
{\em Europhys. Lett.} {\bf 43}, 591 (1998).

\bibitem{dilatancy} M. Piccioni, V. Loreto and S. Roux,
{\em Phys. Rev. E} {\bf 61}, 2813 (1999).

\bibitem{memory} A. Barrat, V. Loreto, {\em Europhys. Lett.} (2000).

\bibitem{nicodemi} M. Nicodemi, A. Coniglio, 
{\em Phys. Rev. Lett.} {\bf 82}, 916 (1999).


\bibitem{alberici} See e.g. F. Alberici-Kious,
J.P. Bouchaud, L. Cugliandolo,
P. Doussineau and A. Levelut, Phys. Rev. Lett. 
{\bf 81}, 4987 (1998).
 
\bibitem{kurchan-laloux} J. Kurchan and L. Laloux,
J. Phys. A {\bf 29} (1996) 1929.

\bibitem{dean}
A. Lefevre, D. S. Dean,
{\em Metastable states of a ferromagnet on random thin graphs},
cond-mat/0011265.

\bibitem{baldassarri}
  A. Baldassarri,
  Laurea Thesis, University of Rome {\em La Sapienza}, (1995).

\bibitem{cude}
  L.F. Cugliandolo, D.S. Dean,
  {\it J. Phys. A} {\bf 28}, 4213 (1995).

\bibitem{bamebu}
  A. Barrat, R. Burioni, M. M\'ezard,
  {\it J. Phys. A} {\bf 29}, 1311 (1996).
 
\bibitem{Bou} J-P. Bouchaud: private communication.

\end{thebibliography}
\end{document}